\documentstyle[aps,eqsecnum,tighten,preprint]{revtex}
\begin{document}
\draft
\setcounter{page}{1}
\title{Mode damping in a commensurate monolayer solid}
\author{L. W. Bruch${}^{1}$ and F. Y. Hansen${}^{2}$}
\address{ ${}^{1}$ 
Department of Physics, University of Wisconsin--Madison\\ Madison,
Wisconsin 53706; \\${}^{2}$
Department of Physical Chemistry,\\ Technical University of Denmark, 
FKI$\cdot$206$\cdot$DTU\\DK-2800, Lyngby, Denmark}
\date{}
\maketitle         
\begin{abstract}  
The normal modes of a commensurate monolayer solid may be damped
by mixing with elastic waves of the substrate. This was
shown by B. Hall et al., Phys. 
Rev. B {\bf 32}, 4932 (1985), for  perpendicular 
adsorbate vibrations in the presence of  an isotropic elastic medium.
That work is generalized with  an  elastic continuum theory of the response 
of modes of either parallel or perpendicular polarization 
for a spherical adsorbate
on a hexagonal substrate.  The results are applied to the discussion of 
computer simulations and inelastic atomic scattering experiments for 
adsorbates on graphite.  The extreme  anisotropy of the elastic 
behavior of the graphite leads to quite different wave vector dependence
of the damping for modes polarized perpendicular and parallel to the
substrate. A phenomenological extension of the elasticity
theory of the graphite to include bond-bending energies improves the
description of substrate modes with strong anomalous dispersion and enables
a semi-quantitative account of observed avoided crossings of the adlayer 
perpendicular vibration mode and the substrate Rayleigh mode.
\end{abstract}

\pacs{PACS numbers: 68.45.Kg, 68.35.Ja, 63.20.-e}

\section{Introduction}
In the course of molecular dynamics calculations of the frequency
spectrum of commensurate monolayer solids of  nitrogen adsorbed on 
graphite [\ref{ref:HB95}, \ref{ref:HBT95}], we became aware of
paradoxical phenomena in what  was expected to be the
most ideal and simple regime. At low temperatures the center-of-mass
one-phonon approximation to the intermediate scattering function has
a nearly pure sinusoidal oscillation  over periods of 20 to 50 $ps$; however
the amplitude generally is quite different from that anticipated from
equipartition theory for harmonic oscillators and from a sum-rule
of Hansen and Klein [\ref{ref:HK}]. 
There is some indication in the simulation data that 
the mean-square oscillation
amplitude, averaged over hundreds of picoseconds, is on the scale  
expected for oscillator coordinates.

The only suggestions of such phenomena which we have found in the literature
are a comment by Hansen and Klein [\ref{ref:HK}] that their sum-rule
was satisfied to 10\% except at small wave numbers where spectral peaks were
quite sharp and a comment by Shrimpton and Steele [\ref{ref:SS91}] 
that simulation times much longer  than 400 $ps$ 
would be needed to ensure thermal equilibration of the long-lived
Brillouin-zone-center phonons of commensurate Krypton/graphite. A 
nanosecond time scale is inferred from experiments and modeling of 
the sliding friction of incommensurate inert gas monolayers 
on metal surfaces [\ref{ref:DK}], but the relative 
importance of processes determining such long lifetimes is in 
dispute [\ref{ref:PN}].                             
We have used perturbation theory for the effect of cubic and quartic
anharmonicity in the adatom--substrate interaction  on the lifetime of
the adlayer phonons[\ref{ref:KK61}, \ref{ref:MF62}]. With parameters 
appropriate to  commensurate Krypton/graphite and to a spherical molecule
version of commensurate Nitrogen/graphite, the  estimated zone-center
phonon lifetimes again are on the scale of nanoseconds. Such processes are
not likely to be the dominant ones in determining the lifetimes,
as has been appreciated by Mills and his co-workers [\ref{ref:HMB},
\ref{ref:HM89}].

The principal mechanism which sets the lifetimes of zone-center phonons
in a commensurate monolayer solid is the radiative damping  arising
because  the adlayer normal mode is actually a surface resonance
that overlaps  a continuum of substrate normal modes  
[\ref{ref:HMB}]. In previous modeling of this process, the substrate was 
treated as an isotropic elastic medium. There was good success in explaining
the phonon line-widths (for motions primarily polarized perpendicular
to the substrate surface) of inert gases adsorbed on metals [\ref{ref:HM89}]. 
The time scale for such damping is of the order of picoseconds and the
damping is expected to be larger for a low-density substrate such as graphite
than for  high-density metal substrates. Inelastic helium atomic scattering 
experiments for xenon adsorbed on  graphite [\ref{ref:TV89}] and for 
commensurate and incommensurate krypton monolayers on  
graphite [\ref{ref:CJD92}] give evidence for a strong  mixing of the adlayer 
perpendicular vibration with substrate modes over a range of wave vectors and 
for strong damping of the normal modes at small wavenumbers.  

Here, we develop the elastic substrate theory for 
the case of adsorption on the (0001) ($c$-axis)  surface of a hexagonal 
substrate. This nominally includes the case of the basal plane surface 
of graphite. However, to mimic the strong anomalous dispersion of
the TA$_{\perp}$ branch of the graphite spectrum [\ref{ref:NWS},
\ref{ref:IV94}],  the continuum approximation of 
Komatsu [\ref{ref:KOM55}] and Yoshimori and Kitano [\ref{ref:YK56}] for 
the bond-bending energies is adopted. 

There is a nonzero frequency at small wave numbers for perpendicular motions 
of an incommensurate monolayer, mainly determined by the curvature of
the adsorbate--substrate potential well, and such modes experience both
hybridization with substrate modes and damping [\ref{ref:HMB},\ref{ref:HG91}].   
For a commensurate monolayer solid, 
there is a Brillouin zone-center gap for motions both parallel and 
perpendicular to the substrate surface; the radiative damping mechanism
acts for both polarizations. 
The long wavelength motions of graphite 
parallel to the surface  plane are governed
by isotropic elasticity theory, a simplification relative to the 
situation for the (111) 
surface of face-centered-cubic metals.   The large elastic anisotropy 
of graphite between motions parallel and perpendicular to the $c$-axis 
has the  consequence that the strong radiative damping is 
confined to a much smaller fraction of the adlayer Brillouin zone for
the branch with parallel polarization than for that with perpendicular
polarization.

The organization of this paper is: Section II describes the models of the
interactions and the intrinsic dynamics of the decoupled adlayer and
substrate. Section III contains the formulation of the coupled adlayer and
substrate dynamics. Section IV contains the formal solution 
for the adlayer response functions.  Some special cases are treated in 
Section V and the results for commensurate monolayers on graphite are
presented in Section VI. Section VII contains concluding remarks.
A summary of our experience with
Molecular Dynamics simulations for the zone-center modes of the nearly
harmonic solid on a static substrate is contained in Appendix A.

\section{Interaction model and intrinsic dynamics}
The required components are  models  for the substrate dynamics, 
the adsorbate--adsorbate interactions,  and the adsorbate--substrate coupling.
The substrate dynamics are modeled with an elastic continuum approximation
which enables  a quite explicit treatment at small wave vectors. Apart from 
an approximation for bond-bending energy terms  in the graphite substrate 
[\ref{ref:KOM55},\ref{ref:YK56}],  the  dispersion of the substrate 
normal modes is omitted. 
The adsorbate--adsorbate interactions are taken to be central pair potentials.
For the small wave number modes,  near the Brillouin zone-center gap,
the form of the pair potential is not crucial to the treatment. 
Finally,  Steele's Fourier decomposition of the 
adatom--substrate interaction is used to make a simple parameterization of the
adsorbate--substrate coupling. As discussed in Sec.III.D, these 
choices affect  the calculated wave vector dependence of the dispersion 
and  damping of the vibrational spectra, but we believe the qualitative
features of the results are reliable. 

In this paper, the $z-$axis of the Cartesian coordinate system is taken 
parallel to the $c-$axis of the hexagonal substrate, the equilibrium
configuration of the adsorbed monolayer 
is in the $x-y$ plane, and the wave vector
{\bf q} of the adlayer normal modes is a 2D vector lying in this $x-y$ plane.

\subsection{Intrinsic substrate dynamics}
The hexagonal solid substrate is approximated  
as an elastic continuum  with 5 independent elastic constants. The equations
of motion for substrate displacements with components u$_i$ are, with the  
summation convention for repeated indices, 
\begin{equation}
\rho \ddot{u}_i = c_{iklm} \frac{\partial^2 u_l}{\partial x_k \partial x_m}\, .
\label{eq:fyh1}
\end{equation}
With hexagonal symmetry, there are only a few nonzero elements in
the fourth rank tensor $c_{iklm}$. In the Voigt notation [\ref{ref:BH}], the
subscripts for Cartesian axes are denoted 1 = $xx$, 2 = $yy$, 3 = $zz$, 
4 = $yz$, 5 = $xz$, and 6 = $xy$.
Then  the independent elastic constants are $C_{11} = C_{22}$,
$C_{33}$, $C_{44} = C_{55}$, $C_{66}$, and $C_{13} = C_{23}$. A further
relation derived from the planar isotropy is 
$C_{12} = C_{11} - 2 C_{66}$. 

For graphite, the mass density is $\rho$ = 2.267 gm/cm$^3$ and the
elastic constants are, all in $10^{11}$ dyn/cm$^2$, $C_{11} = 106$,
$C_{33} = 3.65$, $C_{13} = 1.50$, $C_{44} = 0.40$, and $C_{66} = 44$,
from reference [\ref{ref:RAW}].  Values for $C_{44}$
in the recent literature range from 0.325 to 0.47, derived from 
measurements on highly oriented pyrolytic graphite
with Brillouin scattering [\ref{ref:LL90}] and inelastic neutron 
scattering [\ref{ref:IV94}], respectively. We adopt the 
value  $C_{44} = 0.40$ used in [\ref{ref:RAW}], since it is at the middle of 
the range and this choice facilitates comparison with 
lattice dynamical calculations [\ref{ref:RAWX},\ref{ref:RAWK}].
The speed of the Rayleigh wave [\ref{ref:DM76}] is nearly equal
to that of
the TA$_{\perp}$ mode for wavevector  in the $x-y$ plane and polarization
parallel to the $c$-axis,  $\sqrt{C_{44}/\rho}$, 
because of the large elastic anisotropy of the graphite.

For the graphite substrate, it is
essential to include the strong anomalous dispersion 
[\ref{ref:NWS},\ref{ref:IV94}] of the TA$_{\perp}$ mode. This can be 
accomplished with a continuum approximation to the bond-bending 
energy [\ref{ref:KOM55},\ref{ref:YK56}] by adding the following term to the
total substrate energy:
\begin{equation}
\Delta E = (\lambda / 2) \int d^3 r [ \nabla_2^2 u_z 
        + \partial_z \nabla_2 \cdot \text{{\bf u}} ]^2 \, \, ,
\label{eq:KKb}
\end{equation}
where  the subscript `2' on the gradient denotes the $x-y$ components. 
Eq.(\ref{eq:KKb}) has been constructed so that (i) the associated stress 
tensor is symmetric and (ii)  isotropy in the graphite  basal plane is
retained.  Then for  displacements with spatial dependence
\begin{equation}
\text{{\bf u}({\bf R},t)} \propto \exp (\imath \text{{\bf q}} 
        \cdot \text{{\bf R}} ) \, ,
\label{eq:fyh1a}
\end{equation}
and wavevectors {\bf q} in the
$x-y$ plane, the generalized elasticity theory for motions in the
sagittal plane ($SP$) defined by $\hat z$ and {\bf q} has  the
replacement 
\begin{equation}
C_{44} \rightarrow C_{44}(\text{eff}) = C_{44} + \lambda q^2 \, .
\label{eq:KKc}
\end{equation}
For motions with shear horizontal polarization ($SH$), the `bare'
$C_{44}$ is retained and is denoted $C_{44}^{(0)}$ in this paper. 
We set  $\lambda = \rho K^2$  with $K = 5.04 \times 10^{-3}$ 
cm$^2$/sec fitted to the curvature of the TA$_{\perp}$ branch observed by
inelastic neutron scattering experiments [\ref{ref:IV94}].

\subsection{Adatom-adatom interaction and intrinsic adlayer dynamics}
The adlayer consists of atoms or ``spherical molecules" of mass $m$
in a 2D Bravais lattice; for  commensurate monolayers of
krypton or xenon on graphite, it is a triangular lattice. We assume
that the adatoms interact via a central potential $\psi$
and denote by {\bf r} the projections of the equilibrium positions 
{\bf R} onto the $x-y$ plane. The analysis of this subsection is for modes 
polarized in the $x-y$ plane; the dominant interaction for the out-of-plane
motions arises from the adatom-substrate potential treated in Sec.II.C.
The  normal modes with wavevector {\bf q} have the form
\begin{equation}
\text{{\bf w}}_{j_a} = \text{{\bf w(q)}} 
\exp (\imath [\text{{\bf q}}\cdot\text{{\bf r}}_{j_a}   - 
\omega(\text{{\bf q}})t])   \, .
\label{eq:mode1}
\end{equation}
The amplitude and angular frequency are obtained from
the  solutions of the eigenvalue problem
\begin{equation}
m \, \omega (\text{{\bf q}})^2 \, \text{{\bf w}({\bf q})} 
= \text{{\bf D}({\bf q})} \cdot \text{{\bf w}({\bf q}})
\label{eq:mode2}
\end{equation}
for the dynamical matrix {\bf D}({\bf q}) defined by
\begin{equation}
\text{{\bf D}({\bf q})} = \sum_{j_a \neq 0} \nabla \nabla \psi 
\, [1 - \cos (\text{{\bf q}} \cdot \text{{\bf r}}_{j_a} )] \, .
\label{eq:mode3}
\end{equation}
The small $\vert \text{{\bf q}} \vert $ solutions are transverse and longitudinal acoustic
waves with frequency proportional to $q$. 

The commensurate 
monolayer on a static substrate has in-plane motions with angular 
frequencies given by
\begin{equation}
\Omega (\text{{\bf q}})^2 =  \omega_{0 \parallel }^2   
+  \omega (\text{{\bf q}})^2  \, ,
\label{eq:mode4}
\end{equation}
where $ \omega_{0 \parallel }$ is the zone-center gap defined in
Sec.(2.3). Eq.(\ref{eq:mode4}) provides
the basis for the remark at the beginning of this section that the
analysis near the Brillouin zone center only depends weakly on the
form of the pair potential $\psi$. Thus, we adopt the primitive 
Lennard-Jones (12,6) interaction for  $\psi$;
parameters for krypton and xenon are listed in Table I.

\subsection{Adatom-substrate interaction}
In Sec.III.A, we assume that the interaction of the adlayer atoms or ``spherical" 
molecules j$_a$ with substrate atoms J$_s$ is given by a sum of central pair 
potentials:
\begin{equation}
\Phi_{as} = \sum_{j_a, J_s} \phi (\vert \text{{\bf R}}_{j_a} - 
\text{{\bf R}}_{J_s} \vert) \, .
\label{eq:Steele1}
\end{equation}
For the case of a static substrate lattice with planar surface, $\Phi_{as}$
may be transformed following Steele [\ref{ref:Steele}]: 
\begin{equation}
\Phi\vert_{static} = \sum_{j_a} [ V_o (z_{j_a}) + \sum_{\text{\bf g}} 
V_g (z_{j_a}) \exp (\imath \text{{\bf g}} \cdot \text{{\bf r}}_{j_a})]  \, .
\label{eq:Steele2}
\end{equation}
The  notation in Eq.(\ref{eq:Steele2}) follows that of 
Sec.II.B: the $z$-axis is 
perpendicular to the surface and  {\bf r}$_{j_a}$ is the component 
of {\bf R}$_{j_a}$
parallel to the surface. The {\bf g} are the 2D reciprocal lattice
vectors of the substrate surface. This representation of the static
interaction is more general than Eq.(\ref{eq:Steele1}) since it may include
effects of noncentral forces and many-body forces. For the following,
we truncate the {\bf g}-sum at the first shell of reciprocal lattice
vectors.

In static  substrate  models of a commensurate adlayer such as Krypton/graphite 
with one atom of mass  $m$ per  unit 
cell, the zone-center modes  polarized perpendicular and parallel to the 
surface have angular frequencies given by [\ref{ref:LB88}]
\begin{eqnarray}
\omega_{0 \perp }^2 &=&  (1/m) {d^2 \over dz^2} 
[V_0 (z) + 6 V_g (z) ] \vert_{z = z_{eq}}  \nonumber \\
\omega_{0 \parallel }^2 &=& - 3 g^2 V_g (z_{eq}) /m    \, .
\label{eq:Steele3}
\end{eqnarray}
where the  equilibrium overlayer height is denoted $z_{eq}$.
We  use these frequencies to parameterize the dynamic coupling of the
adlayer to the substrate, with a further assumption 
which is made explicit in Sec.III. Some values of the frequencies, 
based on a combination of experimental data and modeling, are listed
in Table I.

\section{Dynamic coupling of  adlayer and substrate}
In principle, one might solve the coupled dynamics of the adlayer and the
substrate using atomistic interaction models for all the constituents.  
Such calculations have been performed 
[\ref{ref:RAWX},\ref{ref:RAWK}] for the normal modes of coupled 
inert gas--graphite slab systems, but not for the 
effective damping in an adlayer response function.
Further,   there is only limited knowledge of the adatom--substrate 
corrugation energy. Therefore, we  develop a formalism sufficiently detailed to 
show the damping phenomenon and yet one in which the substrate dynamics
and the adatom-substrate coupling are treated with a few empirical
parameters.

\subsection{Parametric forms}
We must first examine the relation
between the descriptions in Secs.II.A and II.C.
In the former the substrate was treated as a continuous medium with
a displacement function {\bf u}({\bf r}, z, t);  in the latter
the  atomic discreteness of the substrate was basic to the lateral 
periodicity of the  adatom--substrate potential, 
the amplitudes $V_g$ in Eq.(\ref{eq:Steele2}).  
The formulation for the dynamic coupling of the adlayer and the continuum 
substrate requires a specification of where 
the stress from the adlayer is applied in the substrate.  
We follow Hall et al. [\ref{ref:HMB}] and assume it to be concentrated
on the surface layer of substrate atoms at height $z_0$, displaced slightly 
inward from  the boundary $z=0$ of the elastic continuum. In the
final results,  $z_0$ is taken to be vanishingly small. 
However, an initial distinction between $z = 0$ and $z = z_0$ is 
made  to bypass  complications of $\delta$-function
stresses applied precisely at the edge of the continuum.

We specialize immediately to the case where the oscillatory displacements
of the substrate and the adlayer are represented by
\begin{eqnarray}
\text{{\bf u}}_{J_s}  &=& \text{{\bf u}}(\text{{\bf q}}, z) 
\exp (\imath [ \text{{\bf q}}\cdot \text{{\bf r}}_{J_s} - 
\omega t])  \nonumber \\
\text{{\bf w}}_{j_a} &=& \text{{\bf w}}(\text{{\bf q}}) 
\exp (\imath [ \text{{\bf q}} \cdot \text{{\bf r}}_{j_a} - \omega t]) \, .
\label{eq:mode5}
\end{eqnarray}
The corresponding interaction energy is derived from the 
second order Taylor series expansion for the potential $\Phi_{as}$ of
Eq.(\ref{eq:Steele1}) 
\begin{eqnarray}
\Delta \Phi_{as} &=& \frac{1}{2} \sum_{j_a,J_s} \nabla \nabla \phi  :
\, (\text{{\bf w}}_{j_a} -\text{{\bf u}}_{J_s}) \, 
(\text{{\bf w}}_{j_a} - \text{{\bf u}}_{J_s})^*  \nonumber \\ 
&=& \frac{N}{2} \sum_{J_s} \nabla \nabla \phi_{j_a J_s} :
\, [\text{{\bf w(q)}} -\exp(\imath \text{{\bf q}}\cdot [\text{{\bf r}}_{J_s} 
-\text{{\bf r}}_{j_a}]) \text{{\bf u(q}}, z)] \nonumber \\
& \times &[\text{{\bf w(q)}} -\exp(\imath \text{{\bf q}}\cdot 
[\text{{\bf r}}_{J_s} -\text{{\bf r}}_{j_a}]) \text{{\bf u(q}}, z)]^* \, ,
\label{eq:mode6}
\end{eqnarray}
where $N$ is the total number of adatoms, the $J$-sum is assumed
to be for atoms in the  $z_0$-layer, and Umklapp processes involving
reciprocal lattice vectors of the adlayer are neglected. 
Although conclusions about the dispersion based on  
an atom--atom model for $\Phi_{as}$ have limited
generality,  such a model was used[\ref{ref:RAWX},\ref{ref:RAWK}]
for commensurate inert gas / graphite cases treated in Sec.VI, so that
it is useful to specify the differences in the approach used here.

Eq.(\ref{eq:mode6}) may be reduced using a tensor generalization
of the analysis which gives Eq.(\ref{eq:Steele2}). However, in view of the
several approximations already made which limit the quantitative accuracy
with which the dispersion may be treated, we make one further  simplification
and drop the phase factor[\ref{ref:PF3}] so that 
\begin{equation}
\Delta \Phi_{as} \approx \frac{N}{2} \text{{\bf K}}_0 : \int dz \, \delta(z - z_0)
        [\text{{\bf w(q)}} -\text{{\bf u}}(\text{{\bf q}},z)] 
        [\text{{\bf w(q)}} -\text{{\bf u}}(\text{{\bf q}},z)]^*   \, ,
\label{eq:mode7}        
\end{equation}
with the tensor coupling constant {\bf K}$_0$ given in dyadic form by
\begin{equation}
\text{{\bf K}}_0 = m ( \omega_{0 \perp}^2 {\hat z}{\hat z} 
+ \omega_{0  \parallel}^2 [{\hat x}{\hat x} + {\hat y}{\hat y}]) \, .
\label{eq:mode8}
\end{equation}

Eq.(\ref{eq:mode8}) is a parameterized representation
for the effect of the adatom-substrate interaction 
in terms of the Brillouin zone-center gap frequencies 
discussed in Sec.II.C. It  may  be 
used to represent the coupling for cases 
where $\Phi_{as}$ is not determined as a sum of pair potentials, 
such as commensurate layers on metals. It is also a way to bypass  the
incomplete  understanding of the origin 
of realistic corrugation amplitudes $V_g$ for adsorption on graphite.
Finally, Eq.(\ref{eq:mode8}) enables a technical simplification 
in the calculation. When combined
with the planar elastic isotropy of the hexagonal surface, the 
problem of coupled adlayer and substrate separates into 
$SP$ and $SH$ motions.

\subsection{Equations of motion}
The equation of motion for the adlayer normal coordinate becomes
\begin{equation}
m \,  \omega^2 \, \text{{\bf w(q)}} = \text{{\bf D(q)}} 
\cdot \text{{\bf w(q)}}  
+ \text{{\bf K}}_0 \cdot [\text{{\bf w(q)}} - 
\text{{\bf u}}(\text{{\bf q}},z_0)] \, .
\label{eq:mode9}
\end{equation}
The differential equations for the components of the substrate amplitude 
{\bf u}({\bf q},z$_0$)
are, for {\bf q} parallel to the $x$-axis and 
$A_{tot}/N$  equal to  the area per adatom in the commensurate adlayer,
\begin{eqnarray}
\rho \, \omega^2 \, u_x (q) &=& (C_{11} q^2 - C_{44} \partial_z^2) 
u_x (q) \nonumber \\
  &-& \imath \, q \, (C_{13} + C_{44}) \partial_z u_z (q) \nonumber \\ 
 &+&  (N/A_{tot}) \, \delta(z - z_0) K_{0xx} [u_x (q) - w_x (q)] \nonumber \\
\rho \, \omega^2 \, u_z (q) &=& (C_{44} q^2 - C_{33} \partial_z^2) u_z (q) \nonumber \\
       &-& \imath \, q \, (C_{13} + C_{44}) \partial_z u_x (q) \nonumber \\ 
       &+&  (N/A_{tot}) \, \delta(z - z_0) K_{0zz} [u_z (q) - w_z (q)] \, .
\label{eq:tmode1}
\end{eqnarray}
and
\begin{eqnarray}
\rho \, \omega^2 \, u_y (q)  &=& (C_{66} q^2 - 
C_{44}^{(0)} \partial_z^2) u_y (q) \nonumber \\
        &+&  (N/A_{tot}) \, \delta(z - z_0) K_{0yy} [u_y (q) - w_y (q)]
\label{eq:mode11}
\end{eqnarray}
 
\subsection{Boundary conditions}
As the boundary conditions on the substrate displacement 
function {\bf u(q}, z),  we take  [\ref{ref:HMB}] the
$z = 0$ boundary to be a free surface where the following components 
of the stress tensor vanish:
\begin{equation}
T_{z \beta} \vert_{z = 0} = 0, \, \, \beta = x, y, z \, .
\label{eq:boun1}
\end{equation}
The components of the stress tensor are given by
\begin{equation}
T_{\alpha, \beta} = c_{\alpha \beta k l} \, \partial_k u_l \, ,
\label{eq:boun2}
\end{equation}
using the 4-index form of the elastic constants. For the
displacement function of Eq.(\ref{eq:mode5}), with {\bf q} parallel 
to the $x$-axis and returning to Voigt notation, 
Eqs.(\ref{eq:boun1}) become (with $z = 0$): 
\begin{eqnarray}
\partial_z u_x (q,z) + \imath \, q u_z (q,z) &=& 0 \nonumber  \\
\partial_z u_y (q,z) &=& 0 \nonumber \\
\partial_z u_z (q,z) + \imath q (C_{13} / C_{33} ) u_x (q,z) &=& 0 \, .
\label{eq:boun3z}
\end{eqnarray}

Finally, the theory of the adlayer response involves substrate motions
driven by an 
initial displacement of adlayer atoms. Then, deep in the substrate, 
$z \to  - \infty$, the
disturbance created by the adlayer must decay exponentially or take 
the form of an  `outgoing' wave. This
becomes a requirement that the solutions 
of  Eqs.(\ref{eq:tmode1}) and (\ref{eq:mode11}) for $z < z_0$  
have the form $\exp(- K \vert z \vert)$ or
$\exp(\imath K \vert z \vert)$  with $K > 0$ -- see Sec.IV.B.

\subsection{Comments}
Eqs.(\ref{eq:mode9}) to (\ref{eq:mode11}) generalize the treatment of
Hall et al. [\ref{ref:HMB}] in two ways: the substrate is  an anisotropic
elastic continuum and there are  driving terms  which
arise from the coherent addition of lateral force terms for the commensurate
adlayer. There is an increase in complexity beyond their treatment,  
but, as shown in Sec.V.C, quite simple results are again 
obtained at the Brillouin zone center.

We summarize the approximations that have been
made which have serious consequences for the treatment of the
dependence of the mode damping on the wave vector:

\begin{enumerate}
\item A distinction is  made between the edge of the elastic
continuum at $z = 0$ and the height $z_0$ where the adlayer stress 
is applied. However the limit $z_0 \to  0$ is taken in the
analysis.
\item The  anomalous dispersion of the TA$_{\perp}$ branch 
of the graphite substrate is approximated in the elastic continuum
description by using the effective elastic constant
C$_{44}$(eff) defined in Eq.(\ref{eq:KKc}). This replaces C$_{44}$ in
Eqs.(\ref{eq:tmode1}). 
It leads to a large shift in the wave number where the
Rayleigh wave of the graphite hybridizes with the $\omega_{\perp}$ adlayer 
mode and improves the agreement with the $HAS$ experiments 
[\ref{ref:TV89},\ref{ref:CJD92}].
\item If the dynamical matrix {\bf D(q)} is dropped from the adlayer             
equation of motion, there is only a small effect for small wave numbers.
\item The approximation in dropping certain phase factors to obtain
Eq.(\ref{eq:mode7}) is  accurate at small wave numbers; however it
omits a $q-$dependence of the dynamic coupling of the
adlayer and substrate [\ref{ref:PF3}].
\end{enumerate}

\section{Correlation functions}
The response of the adlayer in the presence of the substrate is
characterized using the time Fourier transform of  correlation 
functions  of displacement amplitudes defined by
\begin{equation}
S_{\alpha \alpha}(q,t) = \langle W_{\alpha} (q,t) 
W_{\alpha} (q,0) \rangle  \, ,
\label{eq:corrf1}
\end{equation}
where $\alpha = x, y, z$. The initial conditions on the displacements are
\begin{eqnarray}
W_{\alpha} (q,t=0) &=& W_{\alpha 0} \nonumber \\
{\dot W}_{\alpha} (q,t=0) &=& 0 \, ,
\label{eq:corrf4}
\end{eqnarray}
with zero for $t < 0$, and the substrate is initially unperturbed and
static. Then the Fourier transform for Eq.(\ref{eq:mode9}) is
generalized to
\begin{equation}
\int_0^{\infty} \exp(\imath \omega t) \ddot{W}_{\alpha}(q,t) \, dt = 
- \omega^2 w_{\alpha}(q,\omega) + \imath \omega W_{\alpha 0} \, ,
\label{eq:corrf5}
\end{equation}
using
\begin{equation}
W_{\alpha} (q,t) = \frac{1}{2 \pi} \int_{-\infty}^{\infty} 
w_{\alpha}(q,\omega) \exp(- \imath \omega t) \, d\omega \, .
\label{eq:corrf3}
\end{equation}
In this and the following sections, the
dependence on wave number $q$ has been omitted from the notation, to reduce
the complexity of the formulae.

\subsection{Green's function solution}
The solution to the  set of coupled dynamical equations posed in 
Secs.III.B and III.C is conveniently 
stated in terms of the values at $z = z' = z_0$ of 
a set of Green's functions
 g$_{\alpha \beta}$(\text{{\bf q}}, $\omega$, z $\vert z'$) satisfying 
the following set of equations [\ref{ref:DM76}]
\begin{equation}
(\text{{\bf L}} \cdot \text{{\bf g}})_{\alpha \beta} = 
\delta_{\alpha \beta} \delta(z - z') \, ,
\label{eq:corrf6}
\end{equation}
where the $3 \times 3$ differential tensor {\bf L} has the following 
nonzero elements
\begin{eqnarray}
L_{xx} &=& \rho \omega^2 - C_{11} q^2 + C_{44} \partial_z^2 \nonumber \\
L_{zz} &=& \rho \omega^2 - C_{44} q^2 + C_{33} \partial_z^2 \nonumber \\
L_{yy} &=& \rho \omega^2 - C_{66} q^2 + C_{44} \partial_z^2 \nonumber \\
L_{zx} &=& L_{xz} = \imath q (C_{13} + C_{44}) \partial_z \, .
\label{eq:corrf6a}
\end{eqnarray}
The boundary conditions at $z = 0$ based on Eqs.(\ref{eq:boun3z}) are, for
$\alpha = x,y,z$,
\begin{eqnarray}
\partial_z g_{x \alpha} + \imath q g_{z \alpha} &=& 0 \nonumber \\
\partial_z g_{y \alpha} &=& 0 \nonumber \\ 
\partial_z g_{z \alpha} + \imath q { {C_{13}} \over {C_{33}} }
g_{x \alpha} &=& 0 \, .
\label{eq:corrf7}
\end{eqnarray}
The problem separates so that the functions g$_{xy}$, g$_{yz}$,
g$_{yx}$, and g$_{zy}$ vanish [\ref{ref:ggr}].

Then with the definitions
\begin{eqnarray}
\lambda_x &=& (N/A_{tot}) m \omega_{0 \parallel }^2  \nonumber \\
\lambda_z &=& (N/A_{tot}) m \omega_{0 \perp }^2   \, ,
\label{eq:corrf9}
\end{eqnarray}
the functions u$_{\alpha}$(z$_0$) are given in terms of the
functions $g_{\alpha \beta} \equiv g_{\alpha \beta}(z_0 \vert z_0)$ by
\begin{eqnarray}
u_x &=& g_{xx} \lambda_x (u_x - w_x) + g_{xz} 
\lambda_z (u_z - w_z) \nonumber \\
u_z &=& g_{zx} \lambda_x (u_x - w_x) + g_{zz} 
\lambda_z (u_z - w_z) \nonumber \\
u_y &=& g_{yy} \lambda_x (u_y - w_y) \, .
\label{eq:corrf12}
\end{eqnarray}
A formal solution for the driving terms in the
adlayer equations of motion is 
\begin{eqnarray}
u_x - w_x &=& b_{11} w_x + b_{12} w_z \nonumber \\
u_z - w_z &=& b_{21} w_x + b_{22} w_z \nonumber \\
u_y - w_y &=& w_y / (g_{yy} \lambda_x  - 1) \, ,
\label{eq:corrf13b}
\end{eqnarray}
where the coefficients $b_{ij}$ are given by
\begin{eqnarray}
b_{11} &=& [g_{zz} \lambda_z  - 1] / W_b \nonumber \\
b_{12} &=&  - g_{xz} \lambda_z / W_b \nonumber \\
b_{21} &=&  - g_{zx} \lambda_x / W_b \nonumber \\
b_{22} &=& [g_{xx} \lambda_x  -1]   / W_b 
\label{eq:corrf13a}
\end{eqnarray}
and  
\begin{equation}
W_b = [g_{xx} \lambda_x  -1] [g_{zz} \lambda_z  - 1] - g_{xz} 
\lambda_z g_{zx} \lambda_x  \, .   
\label{eq:corrf13}
\end{equation}

The adlayer equations of motion then become
\begin{eqnarray}
[\omega^2 - \omega_{\ell}^2 (q) + \omega_{0 \parallel}^2 b_{11}] w_x 
+ \omega_{0 \parallel}^2 b_{12} w_z &=& \imath \omega W_{x0} \nonumber \\
\omega_{0 \perp}^2 b_{21} w_x + [\omega^2 + \omega_{0 \perp}^2 b_{22}] w_z
        &=& \imath \omega W_{z0} 
\label{eq:gfn2}
\end{eqnarray}
and
\begin{equation}
[\omega^2 - \omega_t^2 (q) + 
(\omega_{0 \parallel}^2 /[g_{yy} \lambda_x  - 1]) ] w_y =  
\imath \omega W_{y0} \, .
\label{eq:gfn3}
\end{equation}
The  $\omega_{\ell}$ and $\omega_t$ are the frequencies of 
longitudinal and transverse polarization, respectively, in the intrinsic
adlayer dynamics, Eq.(\ref{eq:mode2}), and the $x-$axis is taken 
to be a high symmetry direction, $\Gamma M$ or $\Gamma K$, of
the adlayer Brillouin zone.

The solutions are used to form
\begin{equation}
S_{\alpha \alpha}(q, \omega) = \vert w_{\alpha} (\omega, q) \vert^2
\label{eq:gfn3a}               
\end{equation}
with initial condition
\begin{equation}
W_{\beta 0}  =   \delta_{\alpha \beta} \, .
\label{eq:gfn4}
\end{equation}
The damping of the adlayer normal modes manifests itself 
as broadened peaks in $S(q,\omega)$ as a function of $\omega$ 
at fixed $q$. Generally,  peaks in  $S(q,\omega)$ may be assigned as 
derived from the intrinsic adlayer frequencies or from the Rayleigh 
wave of the substrate.  As shown in Sec.IV.B, the radiative damping
mechanism operates for sufficiently small $q$.

\subsection{Evaluation of the Green's functions}
The problem separates into analysis of the $SP$ and $SH$ motions with
the coupled $x-z$ equations and  $y$-equation, respectively. 

\subsubsection{Sagittal plane}
The solution is very similar
to one given by Dobrzynski and Maradudin [\ref{ref:DM76}] for the Green's 
function of a hexagonal elastic half space. 
We seek solutions of the homogeneous versions of Eqs.(\ref{eq:corrf6})
which have exponential  dependences on $z$:
\begin{equation}
g_{\beta \delta} \sim \exp (\alpha z)  \, .
\label{eq:boun4}
\end{equation}
With the definitions 
\begin{eqnarray}
\gamma_1 &=& [\rho \, \omega^2 - C_{11} q^2 ]/C_{44} \nonumber \\
\gamma_4 &=& [\rho \, \omega^2 - C_{44} q^2 ]/C_{33} \nonumber  \\
\sigma_1 &=& \gamma_1 + \gamma_4 + q^2 [(C_{13} + C_{44})^2 /C_{33} 
C_{44}]  \nonumber \\ 
\sigma_2 &=& \sqrt{ \sigma_1^2 - 4 \gamma_1 \gamma_4} \, ,
\label{eq:boun7}
\end{eqnarray}
there are two inverse length scales  $\alpha_j$ given by:
\begin{eqnarray}
\alpha_1^2 &=& [- \sigma_1 + \sigma_2 ]/2 \nonumber \\
\alpha_2^2 &=& [- \sigma_1 - \sigma_2 ]/2  \, .
\label{eq:boun8}
\end{eqnarray}

For $z < z'$, the roots of Eq.(\ref{eq:boun8})  are chosen to give 
damped or outgoing waves according to whether $\alpha_i$ is real or
imaginary[\ref{ref:zroot}].  Denote 
the longitudinal acoustic and transverse acoustic modes for wave vector 
in the $x-y$ plane by  LA (SP$_{\parallel}$) and TA$_{\perp}$,  
respectively, and the  LA mode for wave vector along the $z-$axis
by LA$_z$. The corresponding speeds in the long wavelength limit are
\begin{eqnarray}
c_{LA} &=& \sqrt{C_{11}/\rho}         \nonumber \\
c_{TA} &=& \sqrt{C_{44}/\rho}  \nonumber \\
c_{LAz} &=& \sqrt{C_{33}/\rho} \, .
\label{eq:gfn20}
\end{eqnarray}
The choice of roots for Eqs.(\ref{eq:boun8}) is then
\begin{eqnarray}
\alpha_1 &=& - \imath \vert \alpha_1 \vert \,,  \,  \omega > c_{LA}\,  q \, ,
        \nonumber \\
\alpha_1 &=&  \vert \alpha_1 \vert \, , \, \, \, \, \,   \omega < c_{LA} \,  q \,  ,
\label{eq:gfn21}
\end{eqnarray} 
and
\begin{eqnarray}
\alpha_2 &=& - \imath \vert \alpha_2 \vert \, , \,   \omega > c_{TA} \, q \, ,
        \nonumber \\
\alpha_2 &=&  \vert \alpha_2 \vert \, , \, \, \, \, \,  \omega < c_{TA} \, q \,  .
\label{eq:gfn22}
\end{eqnarray}

According to Eqs.(\ref{eq:corrf6}), the Green's functions are coupled 
in pairs (g$_{xx}$ , g$_{zx}$) and (g$_{xz}$, g$_{zz}$). Then, for 
Eq.(\ref{eq:boun4}) we have
\begin{equation}
g_{x \beta} (\alpha_j) = \imath f_j \, g_{z \beta}(\alpha_j) \, ,
\label{eq:boun8a}
\end{equation}
with the proportionality factor $f_j$ defined  by
\begin{equation}
f_j = -  q \alpha_j (C_{13} + C_{44})/[C_{44} (\alpha_j^2 + \gamma_1)] \, .
\label{eq:boun9}
\end{equation}
The solution of the homogeneous form of Eqs.(\ref{eq:corrf6}) 
in the range $z < z'$ is
\begin{eqnarray}
g_{z \beta} &=& a \exp (\alpha_1 z) + b \exp (\alpha_2 z) \nonumber \\
g_{x \beta} &=& \imath a f_1 \exp (\alpha_1 z) + \imath 
b f_2 \exp (\alpha_2 z)  \, ,
\label{eq:boun10}
\end{eqnarray}
and in the range $z' < z < 0$ is
\begin{eqnarray}
g_{z \beta} &=& [A_{+} \exp (\alpha_1 z)+ A_{-} \exp (-\alpha_1 z)]  \nonumber \\
&+& [B_{+} \exp (\alpha_2 z) + B_{-} \exp (-\alpha_2 z)]  \nonumber  \\
g_{x \beta} &=&  \imath f_1 \, [A_{+} \exp (\alpha_1 z)- \imath 
A_{-} \exp (-\alpha_1 z)]  \nonumber \\
&+& \imath f_2 \,[B_{+} \exp (\alpha_2 z) 
- \imath B_{-} \exp (-\alpha_2 z)]  \, .
\label{eq:boun12}
\end{eqnarray}
The six coefficients $a, b, A_{+}, A_{-}, B_{+}$ and $B_{-}$ are
determined from six equations:
the $z = 0$ boundary condition Eqs.(\ref{eq:corrf7}), 
the continuity of $g_{x \beta}$ and $g_{z \beta}$ at $z = z'$,
and the matching of the discontinuities in the 
first derivatives at $z = z'$ to the strengths of the $\delta$-functions. 
The latter equations are
\begin{eqnarray}
C_{44} \,  [\partial_z g_{xx} \vert_{z=z' +} - 
\partial_z g_{xx} \vert_{z=z' -} ] &=& 1  \nonumber  \\
C_{33} \, [\partial_z g_{zz} \vert_{z=z'+} - 
\partial_z g_{zz} \vert_{z=z' -} ] &=& 1  \, .
\label{eq:boun12b}
\end{eqnarray}
The $z-$derivatives of g$_{xz}$ and g$_{zx}$ are continuous at 
$z=z'$. Completion of the explicit solution for the Green's functions then
is an exercise in linear algebra.

The solutions for g$_{\alpha \beta}(z'\vert z')$ with $z' \to 0$ can be given
in compact form using the definitions:
\begin{eqnarray}
a_{11} &=& q + \alpha_1 f_1 \nonumber \\
a_{12} &=& q + \alpha_2 f_2 \nonumber \\
a_{21} &=& \alpha_1 - q (C_{13} /C_{33} ) f_1 \nonumber \\
a_{22} &=&  \alpha_2 - q (C_{13} /C_{33} ) f_2 
\label{eq:coup4}
\end{eqnarray}
and the Wronskian
\begin{equation}
W_a = a_{11} a_{22} - a_{12} a_{21} \, .
\label{eq:coup4a}
\end{equation}
The Green's function components are
\begin{eqnarray}                    
g_{zx} &=& - \imath (a_{21} - a_{22})/W_a  C_{44}  \nonumber  \\
g_{xx} &=& (f_2 a_{21} - f_1 a_{22})/W_a C_{44}  \nonumber  \\
g_{xz} &=& -\imath (f_2 a_{11} - f_1 a_{12})/W_a C_{33}  \nonumber  \\
g_{zz} &=& (a_{12} - a_{11})/W_a C_{33} \, .
\label{eq:gfn6d}
\end{eqnarray}

There are three  $q-$ranges: I, $q < \omega /c_{LA}$,  where both 
transverse and longitudinal substrate waves are involved in the damping; 
II, $\omega / c_{LA} < q < \omega / c_{TA}$, where only the
transverse waves are involved; and III, $\omega / c_{TA} < q$, 
where there is no radiative damping. Characteristic values for
$\omega$ are  $\omega_{0 \parallel}$ for the parallel
polarization mode and $\omega_{0 \perp}$ for the perpendicular polarization.

\subsubsection{$SH$ mode}
Define
\begin{equation}
\alpha_3^2 = (C_{66} q^2 - \rho \omega^2 )/C_{44}^{(0)} \, .
\label{eq:boun13}
\end{equation}
The speed of the transverse elastic waves in the $x-y$ plane, denoted the  
$SH$ mode, is 
\begin{equation}
c_{SH} = \sqrt{C_{66}/\rho} \, ,
\label{eq:gfn14}
\end{equation}
and the choice of root of Eq.(\ref{eq:boun13}) is
\begin{eqnarray}
\alpha_3 &=& - \imath \vert \alpha_3 \vert \, , \, \omega > c_{SH}\,  q \, ,
        \nonumber \\
\alpha_3 &=&  \vert \alpha_3 \vert \,, \,  \, \, \, \,  
\omega < c_{SH}\,  q \,  .
\label{eq:gfn15}
\end{eqnarray}
Then the solution for $g_{yy}$  has the form[\ref{ref:string}]
\begin{eqnarray}
g_{yy} &=& Y_1 \exp (\alpha_3 z), \, \, \, \, z < z' \nonumber \\
&=& Y_2 \cosh(\alpha_3 z), \, \, z' < z < 0 \, ,
\label{eq:boun14}
\end{eqnarray}
where the $z = 0$ boundary condition, Eq.(\ref{eq:corrf7}), has been used.
The coefficients $Y_1$ and $Y_2$ are obtained from the continuity 
conditions at $z = z'$
\begin{eqnarray}
g_{yy}(z=z'+) - g_{yy}(z=z'-) &=& 0 \nonumber \\
C_{44}^{(0)} \, [\partial_z g_{yy} \vert_{z=z'+} - 
\partial_z g_{yy} \vert_{z=z' -} ] &=& 1  \, .
\label{eq:gfn10}
\end{eqnarray}

The solution for $Y_2$ is
\begin{equation}
Y_2 = - \exp(\alpha_3 z') /(\alpha_3 C_{44}^{(0)}) .
\label{eq:gfn11}
\end{equation}
Then, with $z_0 \to 0$, the value of $g_{yy}$ entering in Eq.(\ref{eq:gfn3})
is
\begin{equation}
g_{yy} = -1/(\alpha_3 C_{44}^{(0)}) \, .
\label{eq:gfn12}
\end{equation}

\section{Special cases}
We discuss here  three special cases where the present
formalism overlaps with other work.

\subsection{Rayleigh wave}
The  frequency (speed) of the Rayleigh wave of wave number $q$ is the 
root of $W_a = 0$, for the Wronskian  defined in Eq.(\ref{eq:coup4a}). 
In the limit  
$q \to 0$, this reproduces the result of Dobrzynski and 
Maradudin [\ref{ref:DM76}].  As noted by others [\ref{ref:RAW}],
in the small$-q$ limit the speed of the Rayleigh wave of the
graphite basal plane surface is only 0.02\% smaller than $c_{TA}$. The
solution for the Rayleigh wave frequency at finite $q$ with the
effective elastic constant Eq.(\ref{eq:KKc}) is formally the same,
but the quantitative results change somewhat. With the parameters
used here, the Rayleigh frequency is 0.1\% smaller than the 
TA$_{\perp}$ frequency at $q=0.3$ {\AA}$^{-1}$ and  0.8\% smaller
at $q=0.6$ {\AA}$^{-1}$.  Even so, the difference between the Rayleigh 
frequency and the TA$_{\perp}$ frequency remains much smaller than 
the 8\% difference found for the case of a Cauchy isotropic elastic 
solid.

\subsection{Isotropic elastic medium}
The formalism reduces to the case treated by Hall et al. [\ref{ref:HMB}]
by choosing
\begin{eqnarray}
\omega_{0 \parallel} &=& 0 \nonumber \\
C_{11} &=& C_{33} \nonumber \\
C_{44} &=& C_{66} \nonumber \\
C_{13} &=& C_{12} \, ,
\label{eq:isot1}
\end{eqnarray}
and examining the structure of the response function 
$S_{ZZ}(q,\omega)$ for
fixed $q$.
Results for the peak frequencies and full-widths at half-maximum for
the damped peaks of $S_{ZZ} (q,\omega)$ are shown  
for a model of Xe/Ag(111) in Figure 1. 
For Figure 1, we extended slightly  the original calculation of Hall 
et al. [\ref{ref:HMB},\ref{ref:fac15}] using 
the parameters $\omega_{0 \perp} = 2.8$ meV  (0.67$_6$ THz) and 
$\rho = 10.635$ gm/cm$^3$ and omitting adatom -- adatom interactions 
($\psi =0$). The effective elastic 
constants C$_{11} = 17.7$ and C$_{66} = 2.86$
(10$^{11}$ dyn/cm$^2$) were fitted to the calculated speeds of longitudinal
and transverse sound for the Ag(111) surface [\ref{ref:HTW}].

Qualitatively [\ref{ref:HMB}, \ref{ref:HM89}], the peak frequencies 
follow trajectories characteristic of an avoided level crossing of the  
substrate Rayleigh wave and the $\omega_{\perp}$ adlayer 
mode at $q \approx 0.3$ {\AA}$^{-1}$. 
For $q < \omega_{\perp} /c_{TA}$, the  $\omega_{\perp}-$mode is damped 
and there is a sharp resonance at a
frequency somewhat reduced (the avoided crossing phenomenon) from that of the
bare Rayleigh wave. At  $q \approx \omega_{\perp} /c_{LA}$, near
0.1 {\AA}$^{-1}$,
there is an additional contribution to the damping and a perturbation
to the peak frequency derived from $\omega_{\perp}$, a  
phenomenon that has been termed a van Hove anomaly [\ref{ref:Zepp90}].
The branch which is the Rayleigh mode at small $q$ 
approaches $\omega_{0\perp}$ at large $q$, but is still 7.5\% below
that limit at 0.4 {\AA}$^{-1}$. 

A novel feature occurs for the present choice of parameters:
there is only one sharp resonance at small $q$, but at
sufficiently large $q$ there are two sharp resonances. One is derived from
the Rayleigh mode and one from $\omega_{\perp}$. The 
second  sharp resonance arises because the upper `repelled' frequency 
lies between the bare substrate Rayleigh frequency  $c_R \, q$ and the 
continuum of substrate frequencies that begins at $c_{TA} \,  q$. 
There is a 6\% difference between  $c_R$ and $c_{TA}$ in this model. 
That there is a signature of the substrate Rayleigh wave at wave numbers
both above and below the  avoided crossing has been considered a notable 
phenomenon in helium atom scattering from adsorbed 
monolayers [\ref{ref:water}]. We do not find the 
corresponding effect in the calculations for  adsorbates on graphite,  
Sec.VI, apparently  because there the increment between the Rayleigh 
frequency and the bulk continuum is quite small.

\subsection{Small q-limit}
In the $q \to 0$ limit, the results of Sec.IV have simple 
explicit forms. The coefficients $g_{zx}$, $g_{xz}$, $b_{12}$, and 
$b_{21}$  vanish, so that the
$w_x$, $w_y$, and $w_z$ motions are decoupled. The  remaining
Green's function components become  (for $\omega > 0$)
\begin{eqnarray}
g_{xx} &=& -1/[C_{44} \alpha_2] 
=   -  \imath/ [\rho c_{TA} \omega] \nonumber \\
g_{zz} &=& -1/[C_{33} \alpha_1] 
=   -  \imath /[\rho  c_{LAz} \omega] \nonumber \\
g_{yy} &=& g_{xx} \, .
\label{eq:smq1}
\end{eqnarray}
Then, defining 
\begin{eqnarray}
\Gamma_x &=&  \lambda_x /[\rho c_{TA} ] \nonumber \\
\Gamma_z &=&  \lambda_z / [\rho c_{LAz}] \, .
\label{eq:smq2}
\end{eqnarray}
the spectral functions are
\begin{eqnarray}
S_{XX}(0, \omega) &=& { {(\omega^2 + \Gamma_x^2)^2} \over
{\omega^2 (\omega^2 + \Gamma_x^2 - \omega_{0\parallel}^2)^2 +
\omega_{0\parallel}^4 \Gamma_x^2}} \nonumber \\
S_{ZZ}(0, \omega) &=& { {(\omega^2 + \Gamma_z^2)^2} \over
{\omega^2 (\omega^2 + \Gamma_z^2 - \omega_{0\perp}^2)^2 +
\omega_{0\perp}^4 \Gamma_z^2}} \, ,
\label{eq:smq3}
\end{eqnarray}
and  $S_{YY}(0, \omega) = S_{XX}(0, \omega)$. 

Approximate expressions for the full-widths at half-maximum 
for $S_{XX}$ and $S_{ZZ}$,  respectively, are
\begin{eqnarray}
\delta \omega_x &\simeq& \Gamma_x \nonumber \\
\delta \omega_z &\simeq& \Gamma_z  \, .
\label{eq:smq5}
\end{eqnarray}
Eqs.(\ref{eq:smq2}) show that the damping
is enhanced for lower density substrates if the other parameters remain 
similar. This indeed is the trend found in comparing the damping of the
$\omega_{\perp}-$modes of Xe/Ag(111) and Xe/graphite, see Sec.VI. 
Eqs.(\ref{eq:smq5})  are accurate for weak damping;  the
results reported in Sec.VI  are obtained with the full formalism of 
Sec.IV and include  self-consistent solutions for cases with strong damping.
                                                                 
Using the $N_2$/graphite parameters in Table I, 
we find $\Gamma_x /\omega_{\parallel} \simeq 0.25$ and 
an estimate of 3 $ps$ for  the decay time. This  supports the
assertion  in Sec.I that the radiative damping mechanism is the
dominant process determining the lifetime of the zone-center mode.

\section{Commensurate monolayers on graphite}
We present applications of the elastic substrate theory of radiative
damping to commensurate  monolayers of Xe/graphite and Kr/graphite
and also compare to the inert gas / 
graphite slab frequency spectra calculated by 
DeWette et al.[\ref{ref:RAWX},\ref{ref:RAWK}].
Although the lateral interactions are the same as in that work, we
have adjusted the frequencies $\omega_{0\parallel}$ and $\omega_{0\perp}$
to incorporate more recent information, so that there are quantitative
differences which arise from differences in the interaction models.

Figure 2 shows the results for the Kr/graphite $\sqrt{3}-$commensurate
monolayer and Figure 3 shows the results for the corresponding Xe/graphite
case. The direction of $q$ is along the $\Gamma K$ axis of the
adlayer Brillouin zone. 
The $\omega_{0\perp}-$frequency is marked as a solid horizontal line in 
both graphs and dotted and dot-dash lines denote the thresholds of 
bulk graphite continua based on the SP$_{\parallel}$ and TA$_{\perp}$ 
modes, respectively. Lateral interactions, with the parameters of 
Table I,
are included following the discussion of Secs.II.B and IV.A.
The plotted points are the  derived peak frequencies of $S_{XX}$ and 
$S_{ZZ}$, as noted, with widths of damped peaks indicated by error bars.
Cases where an error bar coincides with a substrate threshold denote 
a local minimum for the response function, without a full
decrease to half the peak height.

Experiments with Helium Atomic Scattering ($HAS$) [\ref{ref:TV89}, 
\ref{ref:CJD92}] indicate there is a strong damping of the 
$\omega_{\perp}-$mode at small $q$ and show a strong perturbation for
$q \approx 0.25 - 0.3$ {\AA}$^{-1}$   where the TA$_{\perp}-$mode of the 
bare graphite  crosses $\omega_{0 \perp}$. The
extrapolated crossing using the initial slope of the TA$_{\perp}-$mode
is $q \sim 0.4-0.5$ {\AA}$^{-1}$. However, including the strong anomalous 
dispersion of the TA$_{\perp}-$branch with the prescription in 
Eq.(\ref{eq:KKc}), leads to   a semi-quantitative 
account of the crossing. 

Second, we
examine the radiative damping for the in-plane zone-center gap 
$\omega_{0 \parallel}$. The region of strong damping for peaks
of $S_{XX}(q, \omega)$ is
confined to region I defined in Sec.IV.B.1, i.e., to the
left of the substrate SP$_{\parallel}$ threshold shown in the
Figures.  In region II, between the  SP$_{\parallel}$ and TA$_{\perp}$
thresholds, the widths of the peaks in $S_{XX}$ are small but finite and
correspond to lifetimes on the scale of $ns$.
The large elastic anisotropy
of the graphite makes region I much smaller than  for the
isotropic elastic medium: using the parameters for Ag(111) in Sec.V.B
the ratio $c_{TA}/c_{LA}$ is 0.38, but for graphite it is less than 0.1 for 
$q < 0.3$ {\AA}$^{-1}$. Another
manifestation of the large elastic anisotropy is that the elliptical
polarization of the Rayleigh wave nearly degenerates to a transverse
($z$) polarization, with only weak coupling to in-plane motions of the
adlayer. 

Third, the  damping of the $\omega_{\perp}$ branch is very strong 
in both regions I and II. In contrast to the model for Xe/Ag(111), 
Figure 1, we obtain only one sharp resonance (denoted by $+$) 
in $S_{ZZ}$ for region
III. The Figures show a large shift of the resonant
frequency in region III relative to the value for the static substrate 
used as input to the calculation. 

Finally, we  compare the resonant frequencies themselves
with the atomistic calculations of  DeWette and coworkers [\ref{ref:RAWX},
\ref{ref:RAWK}] at large $q$ for a test of the size of
effects of the neglected spatial dispersion. The 
most significant discrepancy is
that `deflection' of the trajectories in the region of the avoided level
crossing  is much larger for the continuum calculation than in the
atomistic calculation. 
In the elastic continuum theory, the 
shift remains on the order of 10\% to the largest $q$ of the
calculation; this is a 50\% larger shift than in the atomistic calculations.
The dispersion with $q$ of the peak `$\omega_x$' ($x$ and $\circ$) 
of $S_{XX}$,  from 
effects of adatom--adatom interactions, is similar to that in
the atomistic calculations. The present
calculations show the $\omega_x$ branch crossing the TA$_{\perp}$ 
branch  (actually, the Rayleigh mode), as do the atomistic calculations.

The main previous test of the elastic continuum theory of radiative
damping was for the damping of incommensurate
inert gas adlayers on Pt(111) [\ref{ref:HM89}], where  
the formalism
tended to underestimate the zone-center damping. The factor of
ten in the mass density of the substrate between graphite and platinum
has the effect of making the damping much larger
for  Xe/graphite. This was anticipated by Toennies
and Vollmer [\ref{ref:TV89}] in their  discussion of  the rather broad peaks 
for the $\omega_{\perp}-$mode   in the $HAS$ inelastic
scattering experiments.

\section{Prospects}
These calculations show that the radiative damping mechanism proposed
by Hall et al. has a major effect on line-widths which may be observed
in inelastic scattering experiments from commensurate adlayers on graphite.
The understanding of the coupling of the commensurate layer to the
substrate is more advanced for substrates such as graphite than for
metallic substrates. Thus, adsorbates on graphite
may be good subjects  for detailed further study. 
It would be of interest to determine whether there are related
effects of substrate dynamics on the monolayer fluid which would disrupt the
{\it long-time} tails seen in molecular dynamics simulations of the
N$_2$/graphite fluid for 1 to 10 ps.

Another question is how to relate the size of the avoided level
crossing of an adlayer mode and the substrate Rayleigh mode
to adlayer--substrate coupling constants. Comparison of our
elastic continuum results to the model calculations of DeWette and
co-workers indicates that there are significant effects of
spatial dispersion to be included. This might be explored in 
future work based on a technique such as lattice Green's 
functions[\ref{ref:HM89}],
now that the elastic continuum theory is in place. For the damping at
intermediate and large wave numbers, where the radiative damping 
mechanism becomes small, a treatment of the more conventional 
anharmonic damping will be needed [\ref{ref:HM89}].

The large differences in the  damping of parallel and perpendicular
motions for the commensurate monolayer on graphite seem well-based
and may have ironic consequences. The  $HAS$ experiments for such monolayers 
could have   more prominent inelastic peaks for
the parallel than for the perpendicular motions, in spite of
the role of polarization considerations in the coupling to
the helium atom to the adlayer. 
            
\section*{Acknowledgments}                            
This work has been partially supported by the National Science
Foundation under Grant No. DMR-9423307 (LWB) and by The Danish 
Natural Science Foundation (FYH).
L. W. B.  thanks the Fysisk-Kemisk Institut and the Technical University
of Denmark for hospitality during the period this work was begun.
We thank for Professor C. J. Goebel and 
Professor H. Taub for several helpful comments and suggestions.
            
\appendix

\section{Molecular dynamics treatment of the zone-center mode}
We summarize the considerations for the center-of-mass one-phonon 
approximation to the intermediate scattering function. Such calculations
for a static-substrate model of the commensurate N$_2$/graphite monolayer 
actually were the starting point of the present paper.

Define the collective coordinate
\begin{equation}
X(\text{{\bf q}}, t) = \frac{1}{N} \sum_{j=1}^{N} 
\exp (\imath \text{{\bf q}} \cdot \text{{\bf R}}_j) \, 
\text{{\bf q}} \cdot \delta \text{{\bf R}}_j(t) \, .
\label{eq:scat1}
\end{equation}
The corresponding intermediate scattering function  is
\begin{equation}
F_1 (\text{{\bf q}},t) = \langle X(\text{{\bf q}}, t) X(\text{{\bf q}}, 0)  \rangle \, ,
\label{eq:scat2}
\end{equation}
with Fourier transform
\begin{equation}
S_1 (\text{{\bf q}}, \omega) = \frac{1}{2 \pi} 
\int_{- \infty}^{\infty} F_1 (\text{{\bf q}},t) \exp (-\imath \omega t) \, dt \, .
\label{eq:scat3}
\end{equation}
This definition omits a Debye-Waller factor present in the analysis
of Maradudin and Fein [\ref{ref:MF62}]. The brackets denote a thermal average
at temperature $T$. 
Hansen and Klein [\ref{ref:HK}]
derived a sum-rule for the classical-mechanics limit pertaining to the
molecular dynamics calculation:
\begin{equation}
\int_{- \infty}^{\infty} \omega^2 S_1 (\text{{\bf q}}, \omega) \, 
d\omega =  k_B T q^2 /N m \, .
\label{eq:scat4}
\end{equation}

In the molecular dynamics calculations at a reciprocal lattice vector   
$\text{{\bf q}} = \mbox{\boldmath$\tau$}$, corresponding to 
the Brillouin  zone-center ``0", there are long 
time-series where $F_1$ appears to have a purely harmonic variation:
\begin{equation}
F_1 (\text{{\bf q}},t) \simeq A_F \cos (\omega_0 t) \, .
\label{eq:scat5}
\end{equation}
Then the sum-rule gives
\begin{equation}
A_F = k_B T \tau^2 /N m \omega_{0}^{2} \, .
\label{eq:scat6}
\end{equation}

Eq.(\ref{eq:scat6}) is  also  obtained from the equipartition of energy
for an oscillator coordinate $Q_{\alpha}$. Take {\boldmath$\tau$} to 
be along an axis $\alpha$: 
\begin{equation}
\frac{1}{2} M \omega_{0}^2 \langle Q_{\alpha}^2 \rangle 
= \frac{1}{2} k_B T \, .
\label{eq:scat7}
\end{equation}
This leads to the same expression for $A_F$, using $M = N m$. The
oscillator representation has the additional consequence that the
fluctuations in $A_F$ may  be derived from the fourth moment of the
Gaussian which leads to Eq.(\ref{eq:scat7}):
\begin{equation}
\langle A_F^2 \rangle = 3 \langle A_F \rangle^2 \, .
\label{eq:scat8}
\end{equation}

Eq.(\ref{eq:scat6}) is badly violated in the results we presented in
Fig.(13) of ref.[\ref{ref:HB95}] for two reciprocal 
lattice vectors of the herringbone lattice of 
N$_2$/graphite. 
At $\text{{\bf q}} = 
1.703 {\hat x} \AA^{-1}$,
the amplitude is much less than the equipartition result (N = 224) and at
$\text{{\bf q}} = 2.95 {\hat y} \AA^{-1}$, it is much greater. While 
those results
were obtained for  time series of $\approx$ 40 ps,  extending the time 
series to 400 ps did not improve the agreement with Eq.(\ref{eq:scat6}) 
by  much.  Results for this pair of reciprocal lattice vectors still had
no simple relation to each other, as individual amplitudes  were
much less or much larger than expected from equipartition theory.  
There was some indication that Eq.(\ref{eq:scat8})
was being satisfied.  

The failure of the $MD$ to match to equipartition strength for {\it good}
harmonic low temperature solids was  observed for the complete 
monolayer $\rho = 1.0$ and for the submonolayer $\rho = 0.5$. 
There also were time intervals
where the apparent amplitude $A_F$ `switched' to another value; there 
was a beating pattern in $F_1$ composed of two (and sometimes three) 
closely spaced frequencies. The phenomena were seen
both in the center-of-mass and in the atomic coordinate versions of the
intermediate scattering functions. The time scales were in the range
$50-200$ ps. Application of the theory of radiative damping shows that 
the dynamic substrate has an important role in
achieving thermal equilibration of the zone-center modes already at 
much shorter times.

\newpage

\section*{Figure Captions}
\begin{enumerate}
\item Model of incommensurate Xe/Ag(111). The frequencies (in THz) for peaks 
in $S_{ZZ}(q,\omega)$ are plotted as a function of wave number $q$ 
(in  {\AA}$^{-1}$). The error flags on the symbols $\triangle$ denote the
frequencies at the half-maxima for given $q$. 
The points $+$ denote sharp resonances.
The solid horizontal line denotes the `bare' frequency $\omega_{0\perp}$ 
and the three straight lines with successively smaller slope are the 
dispersion relations for the longitudinal and transverse acoustic modes 
and the Rayleigh wave at the silver surface.
The substrate is modeled as an isotropic elastic
continuum. See Sec.V.B for the parameters used.
\label{fig:XEA}
\item Commensurate Krypton/graphite. The frequencies (in THz) for peaks 
in $S_{ZZ}(q,\omega)$ and  $S_{XX}(q,\omega)$, as noted, are plotted 
as a function of wave number $q$ (in  {\AA}$^{-1}$) directed along the
$\Gamma K$ axis of the adlayer Brillouin zone. The $\triangle$ and $+$
denote frequencies derived from the peaks in $S_{ZZ}$, while $x$,
$\Box$, and $\circ$ denote frequencies derived from peaks in $S_{XX}$.
The error flags denote the frequencies at the half-maxima in the
response functions. The peaks denoted by $+$ and $\circ$ are sharp
resonances. In the $q$-range where there are no error flags on the 
$x$, the widths of the peaks are finite but too narrow to be
visible on the scale of this graph.  The values of 
$\omega_{0\parallel}$ and $\omega_{0\perp}$ are indicated by dashed and
solid horizontal lines respectively. The thresholds of the TA$_{\perp}$
and SP$_{\parallel}$ continua of the graphite are indicated by dot-dash
and dotted lines respectively. For the
parameters of the interaction model see Table I.
\label{fig:KRG}
\item Commensurate Xenon/graphite.  Identifications as in 
Fig.(\ref{fig:KRG}). Note the crossing of the $\omega_x$ and
Rayleigh branches with no apparent deflection.
\label{fig:XEG}
\end{enumerate}

\newpage

\section*{References}
\begin{enumerate}
\item F. Y. Hansen and L. W. Bruch, Phys. Rev. B {\bf 51}, 2515 (1995).
\label{ref:HB95}
\item F. Y. Hansen, L. W. Bruch, and H. Taub, Phys. Rev. B (in press).
\label{ref:HBT95}
\item J. P. Hansen and M. L. Klein, Phys. Rev. B {\bf 13}, 878 (1976).
\label{ref:HK}
\item  N. D. Shrimpton and W. A. Steele, Phys. Rev. B {\bf 44}, 3297 (1991).
\label{ref:SS91}
\item C. Daly and J. Krim, Phys. Rev. Lett. {\bf 76}, 803 (1996), and 
references contained therein.
\label{ref:DK}
\item B. N. J. Persson and A. Nitzan, (to be published)
\label{ref:PN}
\item V. N. Kashcheev and M. A. Krivoglaz, Sov. Phys.--Solid State {\bf 3},
1107 (1961).
\label{ref:KK61}
\item A. A. Maradudin and A. E. Fein, Phys. Rev. {\bf 128}, 2589 (1962).
\label{ref:MF62}
\item B. Hall, D. L. Mills, and J. E. Black, Phys. Rev. B {\bf 32}, 
4932 (1985).
\label{ref:HMB}
\item B. Hall, D. L. Mills, P. Zeppenfeld, K. Kern, U. Becher, and 
G. Comsa, Phys. Rev. B {\bf 40}, 6326 (1989).
\label{ref:HM89}
\item J. P. Toennies and R. Vollmer, Phys. Rev. B {\bf 40}, 3495 (1989).
\label{ref:TV89}
\item J. Cui, D. R. Jung, and R. D. Diehl, Phys. Rev. B {\bf 45}, 
9375 (1992); R. Vollmer, Ph.D. thesis, University of G{\"o}ttingen, 
1991 (unpublished).
\label{ref:CJD92}
\item R. Nicklow, N. Wakabayashi, and H. G. Smith, Phys. Rev. B {\bf 5},
4951 (1972).
\label{ref:NWS}
\item A. S. Ivanov, I. N. Goncharenko, V. A. Somenkov, and M. Braden,
Physica {\bf B 213 {\&} 214}, 1031 (1995); see 
also H. Zabel, W. A. Kamitakahara, 
and R. M. Nicklow, Phys. Rev. B {\bf 26}, 5919 (1982).
\label{ref:IV94}
\item K. Komatsu, J. Phys. Soc. Jpn. {\bf 10}, 346 (1955).
\label{ref:KOM55}
\item A. Yoshimori and Y. Kitano,  J. Phys. Soc. Jpn. {\bf 11}, 352 (1956).
\label{ref:YK56}
\item For other demonstrations of the hybridization see Y. A. Kosevich
and E. S. Syrkin, Phys. Lett. A {\bf 135}, 298 (1989) and 
P. N. M. Hoang and C. Girardet, Phys. Rev. B {\bf 44}, 1209 (1991).
\label{ref:HG91}
\item M. Born and K. Huang, {\em Dynamical Theory of Crystal Lattices}
(Oxford Press, 1968), Sec.III.11.
\label{ref:BH}
\item E. de Rouffignac, G. P. Alldredge, and F. W. de Wette, Phys. 
Rev. B {\bf 23}, 4208 (1981).
\label{ref:RAW}
\item S. A. Lee and S. M. Lindsay, Phys. Stat. Solidi (b) {\bf 57}, 
K83 (1990).
\label{ref:LL90}
\item E. de Rouffignac, G. P. Alldredge, and F. W. de Wette, Phys. 
Rev. B {\bf 24}, 6050 (1981).
\label{ref:RAWX}
\item F. W. de Wette, B. Firey, E. de Rouffignac, and G. P. Alldredge,
Phys. Rev. B {\bf 28}, 4744 (1983).
\label{ref:RAWK}
\item L. Dobrzynski and A. A. Maradudin, Phys. Rev. B {\bf 14}, 2200 (1976);
{\em ibid.} {\bf 15}, 2432 (E) (1977).
\label{ref:DM76}
\item W. A. Steele, {\em Interaction of Gases with Solid Surfaces} 
(Pergamon, Oxford, 1974).
\label{ref:Steele}
\item L. W. Bruch, Phys. Rev. B {\bf 37}, 6658 (1988).
\label{ref:LB88}
\item F. Y. Hansen, V. L. P. Frank, H. Taub, L. W. Bruch, H. J.
Lauter, and J. R. Dennison, Phys. Rev. Lett. {\bf 64}, 764 (1990).
\label{ref:8a}
\item We  made trial studies of the effect of including the
phase factor, in hopes that this would lead to an improved account of the
magnitude of the splitting of the normal mode frequencies found by 
De Wette et al. at the avoided crossing of the $\omega_{\perp}-$mode
and the TA$_{\perp}-$mode of the graphite. The 
q-dependent  coupling of the adlayer and substrate dynamics is then
smaller, but there
are symptoms of inconsistent improvement of the approximation. In particular,
the resonant frequencies in the modified calculation are no longer 
automatically displaced in the sense of the repulsion usually associated 
with an avoided level crossing.
\label{ref:PF3}
\item  When the members of Eqs.(\ref{eq:corrf6}) and (\ref{eq:corrf7}) 
in which these functions enter are written out explicitly, one finds
that $g_{yz}$ and $g_{yx}$ each satisfy homogeneous differential 
equations, while $g_{xy}$ and $g_{zy}$ satisfy 
coupled homogeneous differential equations with nonvanishing determinant
of the coefficient matrix. The boundary conditions are also homogeneous,
no driving terms, and therefore these Green's functions vanish
\label{ref:ggr}
\item  We implicitly assume that $\sigma_2$ is real. This is valid 
for the isotropic elastic medium, but it is not automatically true
for the hexagonal case. For graphite, with the bond-bending term 
included, $\sigma_2$
becomes imaginary for some frequencies when $q > 1.23$ {\AA}$^{-1}$. That,
however, is beyond the range of wave numbers where we apply the elastic
continuum theory.
\label{ref:zroot}
\item The $y-$motion problem is simple enough that one can identify 
and resolve a pathology arising from the $\delta-$function stress model.
If the Green's function is used to  find the time-dependent substrate 
displacement $u_y(q,z_0,t)$ for an initially static
substrate, the anomaly is  that for $t \to 0$,
there is a nonzero velocity ${\dot u}_y (q, z_0, 0+)$. However, this
arises  because the adlayer force has been concentrated at an infinitesimal
mass element.  Distributing the stress  over a small range $\delta z$ removes
the singular behavior of  ${\dot u}_y$. We  have not identified other places
where the pathology influences the theory of the damping, and have maintained
the   $\delta-$function stress model for all the calculations of this paper.
\label{ref:string}
\item There are no qualitative changes when the enhancement factor F = 1.5 
of Hall et al. [\ref{ref:HM89}] is used.  Then, the undamped upper branch 
emerges from the bulk continuum at $q \approx 0.32$ {\AA}$^{-1}$  
rather than near 0.30 {\AA}$^{-1}$. 
\label{ref:fac15}
\item For a discussion of the elastic constants at the Ag(111) surface,
see U. Harten, J. P. Toennies, and Ch. W{\"o}ll, Faraday  Discuss. Chem.
Soc. (London) {\bf 80}, 137 (1985).
\label{ref:HTW}
\item P. Zeppenfeld, U. Becher, K. Kern, R. David, and G. Comsa,
Phys. Rev. B {\bf 41}, 8549 (1990).
\label{ref:Zepp90}
\item L. W. Bruch, A. Glebov, J. P. Toennies, and H. Weiss, J. Chem. 
Phys. {\bf 103}, 5109 (1995) and references contained therein.
\label{ref:water}
\end{enumerate}

\newpage

\begin{table}
\begin{center}
\caption{Interaction parameters and frequencies of perpendicular 
and parallel  adlayer 
motions at the Brillouin zone center for a commensurate  
$\protect{\sqrt{3}}$
lattice on a static graphite substrate; frequencies are in  meV.$^a$}
\bigskip
\begin{tabular}{|c|c|c|c|c|c|} \hline
Case & $\epsilon^b$ & $\sigma^b$& m$^b$ & $\omega_{0\perp}$ 
& $\omega_{0\parallel}$  \\ \hline 
Kr/graphite & 159$^c$ & 3.60$^c$ & 139.2 & 4.1$^d$ & 1.0$_3^e$ \\
Xe/graphite & 228$^f$ & 3.97$^f$ & 218.0 & 3.0$^g$ & 0.57$^f$ \\
N$_2$/graphite & 95$^h$ & 3.7$^h$ & 46.76 & 6.0$^i$ & 1.6$_5^i$  \\ \hline 
\end{tabular}
\end{center}
a) The conversion factor to THz is 1 meV = 0.242 THz.\\
b) Lennard Jones (12,6) $\epsilon$ in Kelvin and $\sigma$ in {\AA}; mass in 
10$^{-24}$ gm/mol. \\
c) parameters from [\ref{ref:RAWK}]. \\
d) from $HAS$ experiment [\ref{ref:CJD92}].\\
e) Value based on model calculation with enhanced corrugation 
[\ref{ref:SS91}].\\
f) from [\ref{ref:RAWX}].\\
g) from $HAS$ experiment [\ref{ref:TV89}]. \\
h) Values for `spherical nitrogen' to match 3D  critical temperature 
and nearest neighbor spacing in  ground state solid. \\
i) from inelastic neutron scattering [\ref{ref:8a}] on 
orientationally ordered herringbone lattice. \\ 

\label{table1}
\end{table}

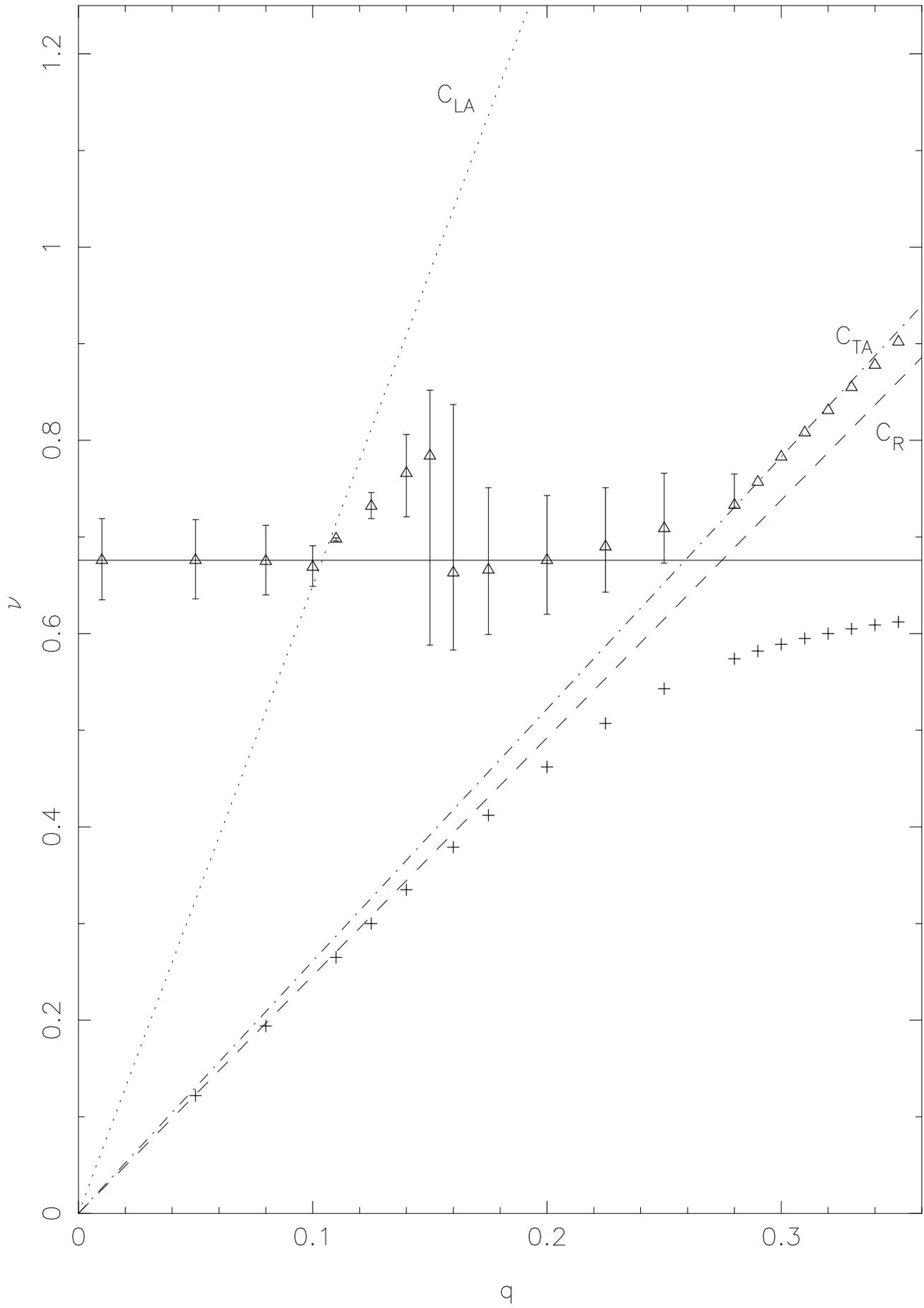
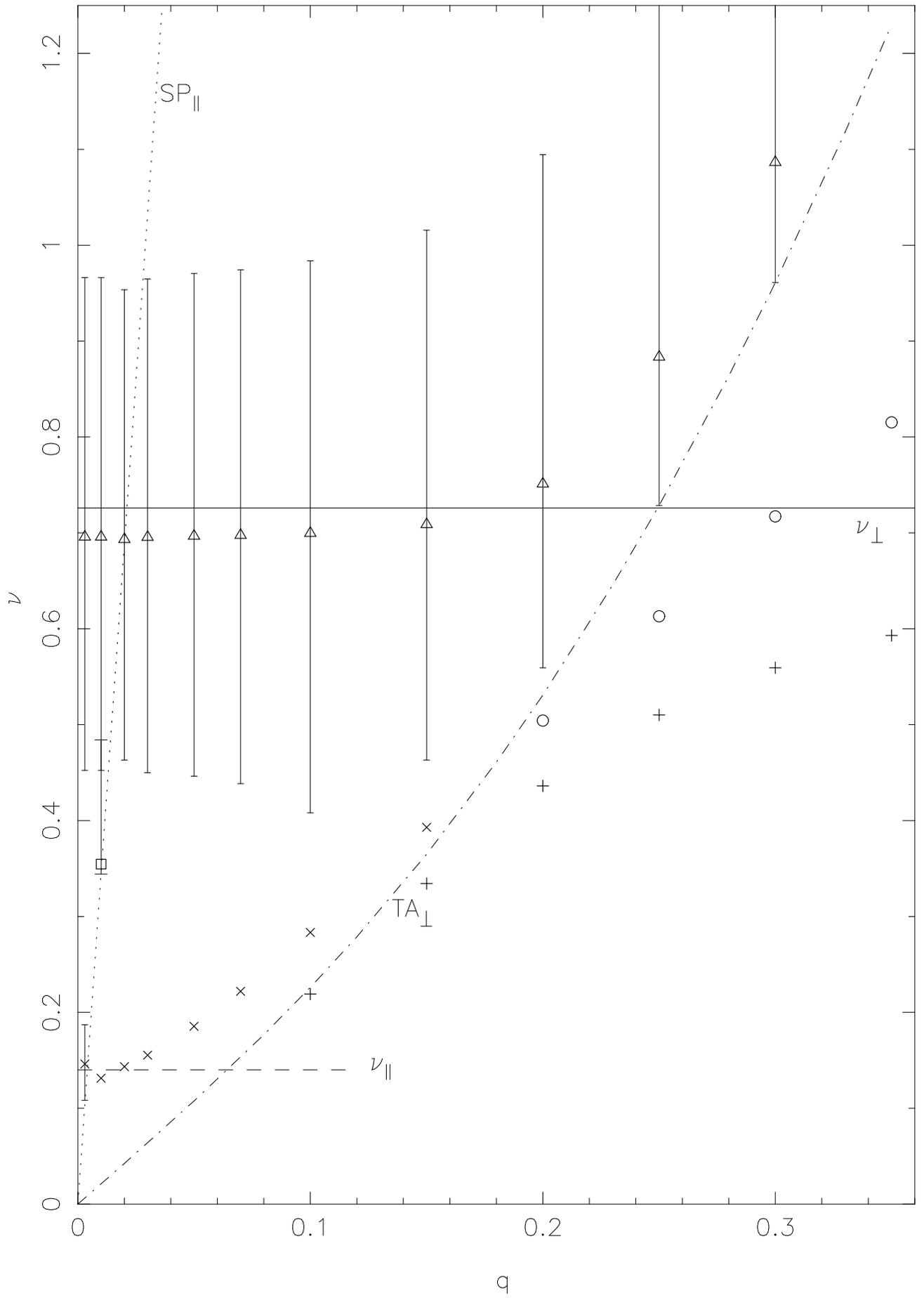
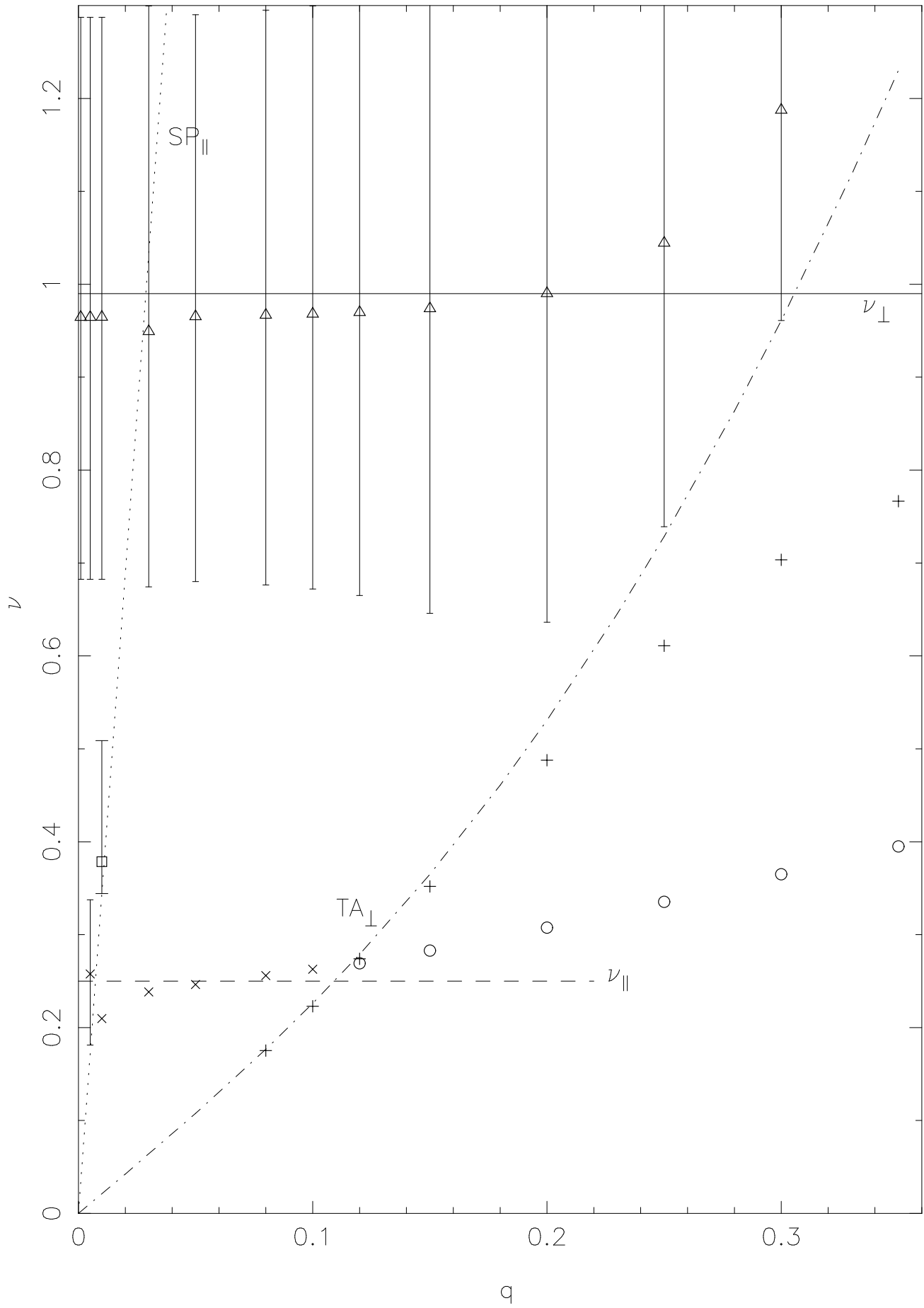
\end{document}